\documentclass[floatfix,preprintnumbers,amsmath,amssymb]{revtex4}

\usepackage{graphicx}
\usepackage{dcolumn}
\usepackage{bm}
\begin{document}
\title{$A=3$ Clustering in Nuclei}
\author{Syed Afsar Abbas}
\email{afsarabbas@yahoo.com}
\author{Shakeb Ahmad}
\email{physics.sh@gmail.com}
\affiliation{Department of Physics, Aligarh Muslim University, Aligarh - 202002, India. }
\date{\today}
\begin{abstract}
Alpha clustering in nuclei, at present is a well studied and reasonably well accepted property of the nucleus. Less well appreciated and more ambiguous is the role of $A=3$ clustering, i.e. helion and triton, in nuclei. Here we try to place $A=3$ clustering in nuclei into its proper perspective, first by pointing out strong experimental evidences which indicate its clear presence in nuclei and secondly showing as to how to include these $A=3$ clusters in a proper and consistent theoretical understanding of the nuclear phenomenon.
\end{abstract}
\maketitle
\section{introduction}
Nuclear physics is studied within the framework of various models. The Independent Particle Shell Model and the Liquid Drop Model form the backbone of current nuclear physics~\cite{a}. However due to strong evidences of the presence of $\alpha$-cluster, (as per current understanding) there has been a growing acceptance of the so called $\alpha$-cluster model.

Here we point out that the equally prominent and equally significant is the role of $A=3$ clusters in nuclei and highlight several experiments which provide equally strong evidences, as that of $\alpha$, of the presence of triton and helion in nuclei. Some of the emprical informations, we feel, are very compelling and should not be ignored. This point of view is however in conflict with the prevailing view that due to the requirement of Pauli Exclusion principle primarily, this just can not be. We examine these arguments in depth and point out as to how to get around this concundrum - that is how to reconcile the compelling evidences of the presence of helion($h$)- and triton($t$)- clusters in nuclei versus the theoretical models which do not really allow them ($h$- and $t$-) as pre-existing entities inside nuclei. These arguments will be shown to lead us to a model which should allow us to attain a better, improved and more internally consistent understanding of the nucleus. 

\section{Experimental evidences of $^{3}\mbox{He}$, $^{3}\mbox{H}$ clustering in nuclei}
Though the concept of alpha cluster studies have been a dominating factor in experimental studies of nuclei, very significant and in some situations, quite prominent, have been experimental studies proving the clear and unambiguos existenece of $^{3}\mbox{He}$ and $^{3}\mbox{H}$ clusters in nuclei. In fact, these empirical evidences of $A=3$ clustering in nuclei are as strong as that of $A=4$ clustering in nuclei.

We agree with Mac Gregor~\cite{b} "{\it .....in order to account for single-particle orbitals in dense nuclear matter, the Pauli exclusion principle must be invoked, and this precludes the existence of any significant nucleon clustering in the interior of the nucleus. But several recent experimental results indicate that extensive nucleon clustering does in fact occur throghout the interior of the nucleus}". He goes ahead in showing that experiments clearly point to evidence of nuclear clustering not only on the surface, but also throghout the interior of the nucleus.

Also ~\cite{b} "{\it one of the most direct ways to establish the existence and the extent of clustering within atomic nuclei is via the direct knockout of particle clusters with high energy projectiles}".

If we are removing $\alpha$ and $t$ and yet leaving the residual nucleus in a state of low excitation, then this situation is impossible if nucleus is only made up of single particle states of shell model. Hence as per this currently prevailing picture there are negligible $t$ and $\alpha$ like structures in the interior of the nucleus. However empirically, if we do find significant amount of $\alpha$- and $t$- coming out of these collisions in knockout processes, then one is forced to conclude that these should have been already pre-existing in there.   

Artun et al.~\cite{c} bombarded 150-, 230-, 600-, 960- MeV protons on several targets of N, O, P, S, V, Al, Si, Ca, and Fe. Prompt $\gamma$ rays were measured which provided information on the removal of $t$- and $\alpha$- from these nuclei when the residual nucleus was left in low excited states.  Clean evidences for the knockout of $\alpha$-, $h$-, $t$- as single entities etc. were found. But the probability of knocking out such complex clusters as single entities while leaving the residual nucleus in low excited states, as stated above, would be extremely low in a shell model. Even if one is willing to accept some kind of formaton of $\alpha$ on the surface (say, due to high binding energy of 28 MeV), it would still be puzzling as to how come only $\sim$8 MeV bound $^{3}\mbox{He}$ and $^{3}\mbox{H}$ are removed as a single block. The only way to understand this result is to accept it as a fact that these were actually always there, in the preformed form, inside these nuclei.

Poskanzer et al.~\cite{d} measured the $^{3}\mbox{He}$ and $^{4}\mbox{He}$ fragments emitted from Ag and U targets when bombarded with 2.7 GeV proton, and 1.05 GeV/nucleon $\alpha$ particles and $^{16}\mbox{O}$ ions. They noticed dramatic increase of cross sections with projectile mass. Their setup had good separation and detectability of $^{3}\mbox{He}$. We quote their conclusion, "{\it ...... our data present evidence for the {\bf nonevoporative} emission of $^{3}\mbox{He}$ and, to a somewhat lesser extent, $^{4}\mbox{He}$ products in collisions between relativistic heavy ions. The cross sections for these high-energy products are two to three orders of magnitude higher than those found for proton-induced reactions at comparable energy. This points towards a cooperative   mechanism that can not be explained by geometrical considerations or by an independent superpositions of nucleon-induced knockout cascades}". To us, this indicates a clearcut evidence, in particular, of already pre-existing $^{3}\mbox{He}$-clusters in those nuclei. Indeed this is such a clean evidence of $^{3}\mbox{He}$ as a well formed cluster inside these heavy nuclei.

Not only these, as pointed out by Mac Gregor~\cite{b}, there are several other experiments pointing to pre-existing $t$-, $h$- and $\alpha$- clusters in nuclei. Existence of alphas could not be too surprising to many at present. However, the existence of $t$ and $h$ as clusters inside nuclei should be puzzling to many. But there they are, as per these experiments!

Next,``{\it The electron induced quasi-elastic nucleon knockout reaction is one of the cleanest probes of the single particle (SP) structure of a nucleus. For light 0p-shell nuclei like $^{6}\mbox{Li}$ this reaction assumes a further dimension: it may reveal the relative importance of SP aspect and clustering aspect~\cite{e}}".

The above is true as the knockout of a single proton from complex nuclei by high-energy electrons is assumed to proceed predominantly via a one-step reaction. Experimental deviations from this impulse-approximation ansatz have been well reported in literatures~\cite{f}. Indeed what has been observed is direct trinucleon - both $^{3}\mbox{H}$ and $^{3}\mbox{He}$, knockout from $^{6}\mbox{Li}$ via exclusive electron reaction~\cite{g}. What they measured was the mirror $^{6}\mbox{Li}(e,e^\prime {^{3}\mbox{He}}){^{3}\mbox{H}}$ and $^{6}\mbox{Li}(e,e^\prime {^{3}\mbox{H}}){^{3}\mbox{He}}$ reactions. The momentum transfer dependence is in complete disagreement with the fundamental expectrum of a direct-single nucleon knockout. Whereas the momentum dependence is in good agreement with a direct $A=3$ knockout mechanism - indicating that these $h$- and $t$- clusters exist as primary entities in $^{6}\mbox{Li}$. They also compared this with the $^{4}\mbox{He}(e,e^\prime {^{3}\mbox{He}})n$ and $^{4}\mbox{He}(e,e^\prime {^{3}\mbox{H}})p$ which in fact gave clear evidences of the existence of $^{3}\mbox{He}$ and $^{3}\mbox{H}$ as pre-existing clusters in $\alpha$-nucleus~\cite{g}.

To belabour the point, for the sake of clearity, if in an exclusive electron scattering we get a direct single nucleon knockout then we are forced to conclude that indeed there was a pre-existing nucleon inside the nucleus which got knocked out in this process. If on the other hand we find out, that inspite of our gut feeling, a single $h$- or $t$- is getting knocked out directly from the nucleus, then the only logical conclusion should be that these were already there in the first place, as pre-existing in the target nucleus. Not only this, they also appear to have a primacy which we do not normally associate to them i.e. $h$ and $t$, but only to $n$ and $p$. But if $n$ and $p$ are primary due to the direct knockout experiment, so should $h$- and $t$- be primary due to the similar above direct knockout experiments.

Recently Ohkubo and Hinabayashi~\cite{h} in the first experiment of its kind, provide evidence for higher nodal band states with $^{3}\mbox{He}$ cluster structure in $^{19}\mbox{Ne}$ and prerainbow in $^{3}\mbox{He}+{^{16}\mbox{O}}$ scattering. We quote them~\cite{h}, "{\it The present findings about the higher nodal band states with the $^{3}\mbox{He}$ cluster in $^{19}\mbox{Ne}$ in addition to the higher nodal states with the $\alpha$-cluster structure to $^{20}\mbox{Ne}$, $^{40}\mbox{Ca}$ and  $^{44}\mbox{Ti}$ nuclei and the higher nodal states with the $^{16}\mbox{O}$ cluster structure in $^{32}\mbox{S}$ nuclei reinforce the importance of the concept of the higher nodal state and the $^{3}\mbox{He}$ cluster in nuclei}". We wish to remark here that the existence of $\alpha$-clusters and $^{3}\mbox{He}$ clusters in adjoining nuclei not only enforces the view that as a cluster $^{3}\mbox{He}$ is as basic as $^{4}\mbox{He}$ is, but also points to the fact that $^{4}\mbox{He}$ itself has a subcluster structure $^{4}\mbox{He}\sim {^{3}\mbox{He}}+n$.

Cunsolo et al.~\cite{i} used the reaction $^{14}\mbox{C}({^{6}\mbox{Li}},t){^{17}\mbox{O}}$ at $E(^{6}\mbox{Li})=34 MeV$. The observed selectivity and the forward peaked angular distribution suggested a predominantly direct reaction mechanism. Here they confirmed a dominant direct $^{3}\mbox{He}$ transfer in this reaction.

The inadequacy of the single-particle shell model to explain the structure of low lying negative parity states of say $^{17}\mbox{O}$ and the multiplicity of positive parity states has been confirmed by many experiments. We quote a few such experiments indicating presence of $^{3}\mbox{He}$ and $^{3}\mbox{H}$ in nuclei~\cite{j}. 

Mac Gregor~\cite{b} has pointed out several experiments which give strong evidence of presence of $\alpha$-, $h$-, and $t$- clusters in nuclei. For those extensive references we refer the reader to Mac Gregor's paper~\cite{b}. These experiments plus the experiments which we have quoted above, inspite of our pre concerned notions, give strong and compelling evidences of the primacy existence of $A=3$ ($h$- and $t$-) clusters (as well as $\alpha$-cluster) in nuclei.

\section{Theory of clustering in light nuclei}
Matching the extensive experimental studies looking for clustering in light nuclei, is the large number of theoretical studies of these nuclei. To set up the framework of most of these studies we quote from the authoritative article of Ikeda et al.~\cite{k}. Because of the significance of these ideas in the framework of our model we quote them extensively and also we post some comments right here.

"{\it In light nuclei, there has been clarified that a structure different in quality from the shell structure-that is the molecule-like structure as a well developed cluster structure - appears systematically in the low-lying states near the ground state......}".

({\bf comment:} clearly cluster formation is not permitted in shell models.)

"{\it Appearance of the molecule-like structure, especially, the one with a fundamental unit of $\alpha$-cluster, indicates that the clustering correlations, especially $\alpha$-clustering correlations, are prominanat correlations in the light nuclear system}".

({\bf comment:} clearly these clustering correlations are not part of the shell model structure. Also if $\alpha$-cluster requires specific $\alpha$-cluster correlations, then the presence of triton clusters should be due to specific three-body correlations which should be different from the other one relevant for the $\alpha$-cluster. The point is that, if say in $^{7}\mbox{Li}$, $\alpha$ and a $t$-cluster are well formed. Then as per the above idea, $\alpha$-cluster arises due to specific four body correlations and in $t$-cluster there should be a different three body correlation leading to the formation of these clusters inside $^{7}\mbox{Li}$. Note that this is clearly against the effect of antisymmetrisation of seven nucleons in $^{7}\mbox{Li}$).

"{\it if we understand the fundamental characters of clustering correlations as to be the factors necessary for the formation of the molecule-like structure, they can be said to be {\bf strong internal cluster correlations and weak inter-cluster correlations}. This expresses firstly that the internal binding of the constituent clusters is strong similarly as in their isolated situation and secondly that the inter-cluster interaction is weak in the region where the saturated clusters preserve their identity approximately. Since these characters of the correlations are generated in real nuclei, they are intimately connected with the action of Pauli principle which gives strong restrictions to nuclear structure. In the case of ground state of usual nuclei which has a spatially compact configuration of nucleons, this pauli principle acts as a healing function to the shell model orbits and brings about the independent-particle aspect. On the other hand in the molecule-like structure, the pauli principle acts as an effective repulsive core force in the overlapping region of two $\alpha$-nuclei and plays a role to sustain the molecule-like structure through strenghthening the characteristics of the clustering correlations. these distinctive functions of the pauli principle in two different kind of structure (or phase) were called as the {\bf dual role of the pauli principle}}". 

({\bf comment:} Hence there is a a one-to-one analogy of this picture with another physical reality of hadronic matter. A multiquark system is confined in QCD as per colour singlet hypothesis. Hence antisymmetrisation occurs for all the quarks in the full $SU(3)_C\times SO(3)\times SU(3)_F\times SU(2)$ space (exactly as in the case of A nucleons in a nucleus in independent particle shell model as indicated above). But this mutiquark breaks up into Pauli antisymmetrised clusters of only three quarks each, to give individual nucleons and which in turn have a weaker resulting nuclear force between them to provide an overall bound nucleus (this is exactly as above "{\it strong internal cluster correlations and weak inter-cluster correlations}"). Indeed analogously as above there is a "{\it dual roles of the Pauli principle}". This point is basic and deep and relates to how the $SU(2)$ isospin of $(u,d)$ quarks changes into the $SU(2)$ isospin at the nucleonic $(n,p)$ level as discussed below.)

"{\it The fact that the molecule-like structure appears generally in the excited states of nuclei means the occurence of structure-change as a kind of phase change of the nuclear many-body system due to the clustering correlations.}"

({\bf comment:} These putative clustering correlations should be there even in these clusters when these are free entities. Also here ground state is believed to be shell model kind while the clustering occurs in the excited state only. This belief is quite prevailent~\cite{l}. There, for example the state in $^{6}\mbox{Li}$, the ground state $\alpha$-particle is single $(0S)^4$ harmonic oscillator state and the first excited state may have the 3N+N structure. (which should read a 3N cluster structure plus N). But this is not correct in more recent theoretical model studies where in the ground state itself, there are prominant clustering effects~\cite{m,n}. Note that when the $\alpha$-particle is 3N+N, then clearly what one is saying is that it is made of a 3N cluster plus a single nucleon. However as per above the "{\it dual role of the Pauili principle}" in $\alpha$, there should be a Pauli principle of two different kind - one that works for $A=3$ cluster and the other for the fourth nucleon with respect to $A=3$ cluster treated as a single basic entity.)

Hodgson has given a recent review of $\alpha$-particle clustering in nuclear reactions~\cite{o}. A few quotations from there and our comments.

"{\it All aspects of nuclear structure and reactions involving clusters can of course be treated in terms of the individual nucleons making up the clusters but this is not only far more complicated but may lose the physical insight obtained by thinking in terms of clusters. It may be remarked that this applies also to the nucleons themselves, as they are believed to be quark clusters}".

({\bf comments:} This point arises basically from the antisymmetrization issue of Ikeda~\cite{k}  and that of different clusters as being equivalent and which in turn is equivalent to the shell model; but as we discuss below, is not correct. Clustering is due to specific correlations which are not part of the shell model structure - see discussion in Ikeda paper above. Clearly clustering is demanding new degrees of freedom in nuclei.)

"{\it of all such clusters, that formed by two protons and two neutrons is the most likely because of high symmetry and binding energy"}.

({\bf comment:} If this were true then how come $A=3$ clusters $t$ and $h$ exist inside nuclei as clearly evident from experimental data. These are only $\sim$8 MeV bound and are neither symmetric. So as such these $A=3$ entities should not exist as clusters inside nuclei, but indeed they actually do. So the reason for this to occur can not be just due to high binding of $\alpha$ and its symmetry, but some another reason has to be sought. We shall show below that the reason why $t$, $h$ and $\alpha$ form clusters inside nuclei is because all the three have unique and similar density profile along with a prominent hole at the center of these nuclei and which is very different from the density profile of all the other nuclei.)

\section{Equivalence of Different multi-cluster structures}
One of the most dominating concept in the theoretical understanding of light nuclei and that of the cluster structure therin, has been the concept of equivalence of different multicluster structures in nuclei~\cite{p}. For example in the Oscillator Model the ground state of $^{6}\mbox{Li}$ is described by the following wave functions of equivalent form~\cite{p}[p.40, AppendixC]
\begin{eqnarray}
\psi_{gs}(^{6}\mbox{Li})&=&A\{\phi_0(\alpha)\phi_0(d)\chi(\alpha-d)\}\\ \nonumber
&=&NA\{\phi_0(t)\phi_0(h)\chi(t-h)\}
\end{eqnarray}
Here $\chi(\alpha-d)$ and $\chi(t-h)$ are two oscillator quanta wave functions and $\phi_0(\alpha)$, $\phi_0(d)$, $\phi_0(t)$ and $\phi_0(h)$ are the internal wave functions of $\alpha-$, $d-$, $t-$, and $h-$ clusters respectively. The constant $N$ is here to ensure that the two functions are equal to each other. The mathematical equivalence of the two can be easuily demonstrated~\cite{p}. Since as per this picture, they are the same, hence either description is as good as the other and also that these two descriptions may coexist simultaneously for $^{6}\mbox{Li}$. And also note that in this picture these two are not orthogonal either. Similar putative similarities for other light nuclear clusters have been dominating in the background in the current studies of other light nuclei~\cite{p}.

Here we give a few examples of the persistent influence of the above ''{\it multi-cluster-equivalence}'' concept in current nuclear physics. In studying triton-cluster knockout for $^{6}\mbox{Li}$ the authors~\cite{g} discussed the surprising predominance of $^{3}\mbox{He}-{^{3}\mbox{H}}$ cluster in $^{6}\mbox{Li}$ ground state vis-a-vis the expected $^{4}\mbox{He-d}$ state, ``{\it It should be pointed out that large values of cluster probabilities for both $^{2}\mbox{H}-{^{4}\mbox{He}}$ and $^{3}\mbox{He}-{^{3}\mbox{H}}$ configurations are not mutually exclusive, because of the corresponding wave functions are not orthogonal~\cite{g}}''. Note that here this putative lack of orthogonality and equivalence is as per the point discussed above.

Another example is from the authoritative review of cluster models of light nuclei, Langanke~\cite{n} in discussing the surprising predominance of $(p+{^{3}\mbox{H}})$ and $(n+{^{3}\mbox{He}})$ components over $(\mbox{d+d})$ configurations for the ground state binding energy of $^{4}\mbox{He}$ nucleus states, tries to patch up this conundrum by stating, "In the extreme Shell Model limit (Harmonic Oscillator wave functions and identical width parameters) antisymmetrized 3N+N and 2N+2N cluster functions becomes identical and represent the same $(0S)^4$ shell model wave functions".

To understand the issue, let us examine the basis for the assumptions of as to the equivalence of $4+2=3+3$ for $^{6}\mbox{Li}$ and $3+1=2+2$ for $^{4}\mbox{He}$ etc. in the currently well accepted picture of light nuclei.

The most significant fundamental assumption in establishing the mathematical equivalence of various clusters in these model is that one {\bf takes into account only two body forces and completely ignore the three body forces}~\cite{p, p.113}. This is one of the basic ansatz in this picture. But what it means is that clearly if one is forced to include (because of empirical reasons) any kind of three body forces, then the above equivalence would be completly lost.

Now, until recent times nuclear physicists were quite content with only using two body forces. But in recent years it has been clear that even for $A=3$ nucleus one just can not do consistent and good nuclear physics without including three body forces. So it is wrong to just have two body force and discard three body force (we discuss this below). Thus the very basis of this putative equivalence of multicluster-cluster in light nuclei, and which as discussed above forms the backbone of practically all analyses, experimental or theoretical, of light nuclear cluster structures and {\bf as shown above is quite incomplete and wrong}. Because of the fundamental requirement of three body forces, the whole concept of the equivalence of various clusters should be discarded. The above spurious concept has prevented people from having a proper understanding of the true nature of clustering in nuclei~\cite{q}. 

\section{Why Three Body Forces are Needed Fundamentally?}
In trying to fit the nuclear matter binding energy and density simultaneously the results of the best two body interactions, all fall on the so called Coester band. No matter what you do, if you confine yourself to two body interaction then there is no way one can do any better. This clearly hints at including atleast a three body force [3BF]. There are also several compelling reasons why it seems that it is not enough to confine ourself to only two body interactions. For example, there are basic phenomenological reasons~\cite{r} why a three body force is needed and also more fundamental reasons~\cite{s} for the same. Note that these 3BF can not be reduced to any two body force and thus have their own independent existenmce.

The realization that in nonrelativistic framework 3BF is indispensable in a complete and consistent description of nuclei, has grown in recent years. One knows that $A=3$ bound states $^{3}\mbox{H}$ and $^{3}\mbox{He}$ are theoretically underbound using only NN (two body) forces. So also is $^{4}\mbox{He}$ underbound and this is represented by the so called Tjon lines [20]. Proper inclusion of 3BF in different models is essential to fit the binding energies of all $^{3}\mbox{H}$, $^{3}\mbox{He}$ and $^{4}\mbox{He}$~\cite{t}. The same paper also indicated that the presence of four-body forces (4BF) is ruled out for $^{4}\mbox{He}$.

This point is very succintly made by Wringa and Peiper~\cite{u} "{\it Modern nucleon nucleon (NN) potentials, such as the Argonne $v_{18}$, CD Bonn, Reid93, NijmI, and NijmII, fit over 4300 elastic NN scattering data with a $\chi^2=1$. These potentials are very complicated, including spin, isospin, tensor, spin--, quadratic momentum-dependent, and chrage-dependenct tensor, with $\sim 40$ parameters adjusted to fit the data. Despite this sophistication, these potentials can not reproduces the binding energy of few body nuclei such as $^{3}\mbox{H}$ and $^{3}\mbox{He}$ without the assistance of a three-nuceon potential}".

Henec we see that the 3BF manifest itself right away in the structure of $A=3$ nuclei, so much so that it is not possible to fit triton's and helion's binding energy, rms radii etc. without involving a 3BF~\cite{t,u}. Note also that 4BF in $\alpha$-nucleus is negligible~\cite{t}.

\section{Why $h$-, $t$-, and $\alpha$- form subclusters inside nucleus?}
It is very often stated that formation of $\alpha$-clusters inside nuclei is due to the fact that $\alpha$- has such large binding energy of 28 MeV~\cite{o}. But then $t$- and $h$- have much smaller binding energy $\sim$8 MeV. Therefore if this was the only requirement for the creation of cluster inside nucleus then $h$- and $t$- clusters should not exist inside nuclei. But as we have seen there have been several experiments which have clearly indicated presence of $t$- clusters and $h$- clusters in nuclei. If the reason for cluster formation inside nuclei is not only due to their binding energies then why do these cluster exist at all inside nuclei?
\begin{figure}[t]
\includegraphics[height=8cm,width=8cm]{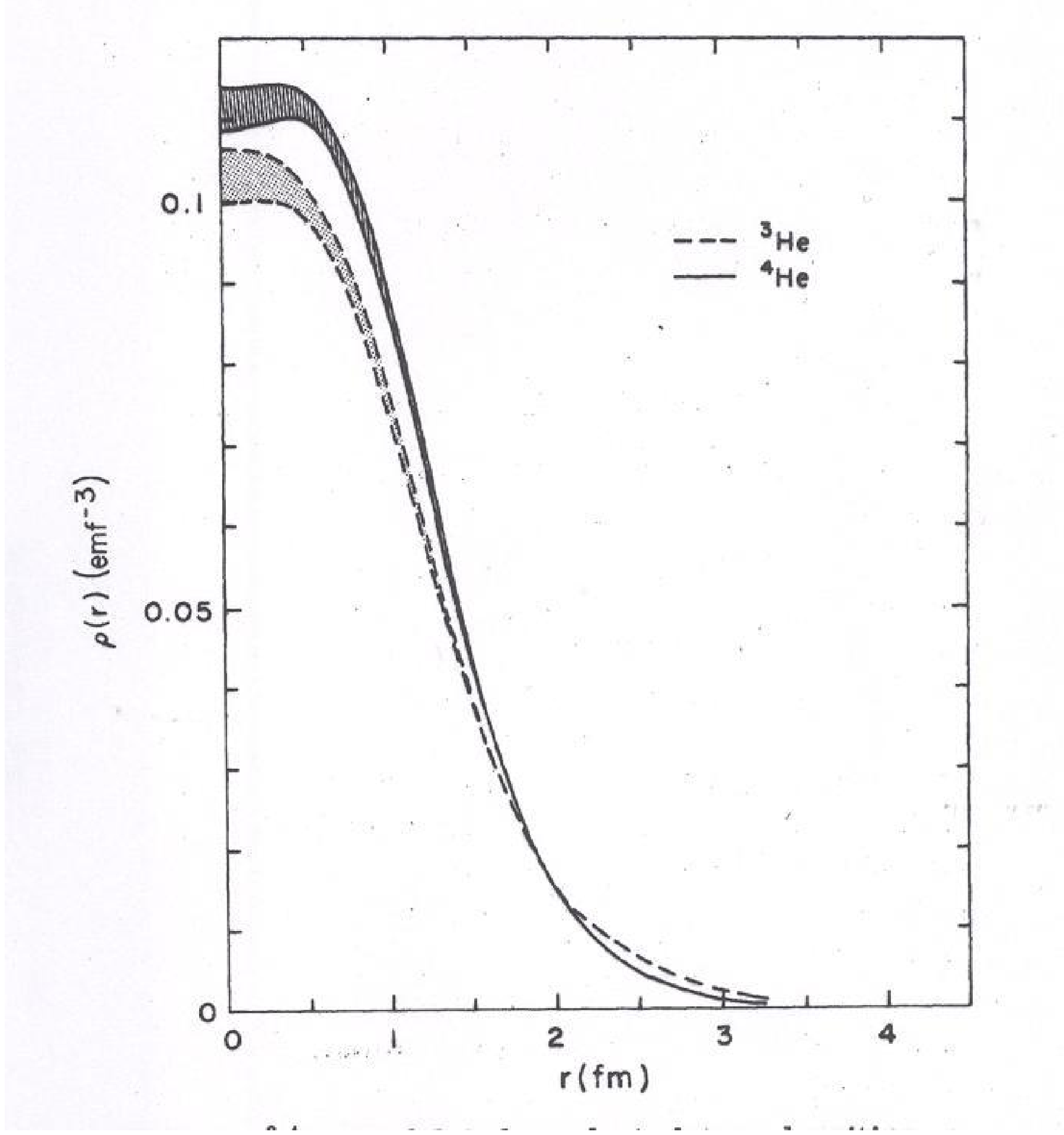}
\caption{Density distribution of $^{3}\mbox{He}$ and $^{4}\mbox{He}$ from ref.~\cite{v}}
\end{figure}
\begin{figure}[t]
\includegraphics[height=8cm,width=8cm]{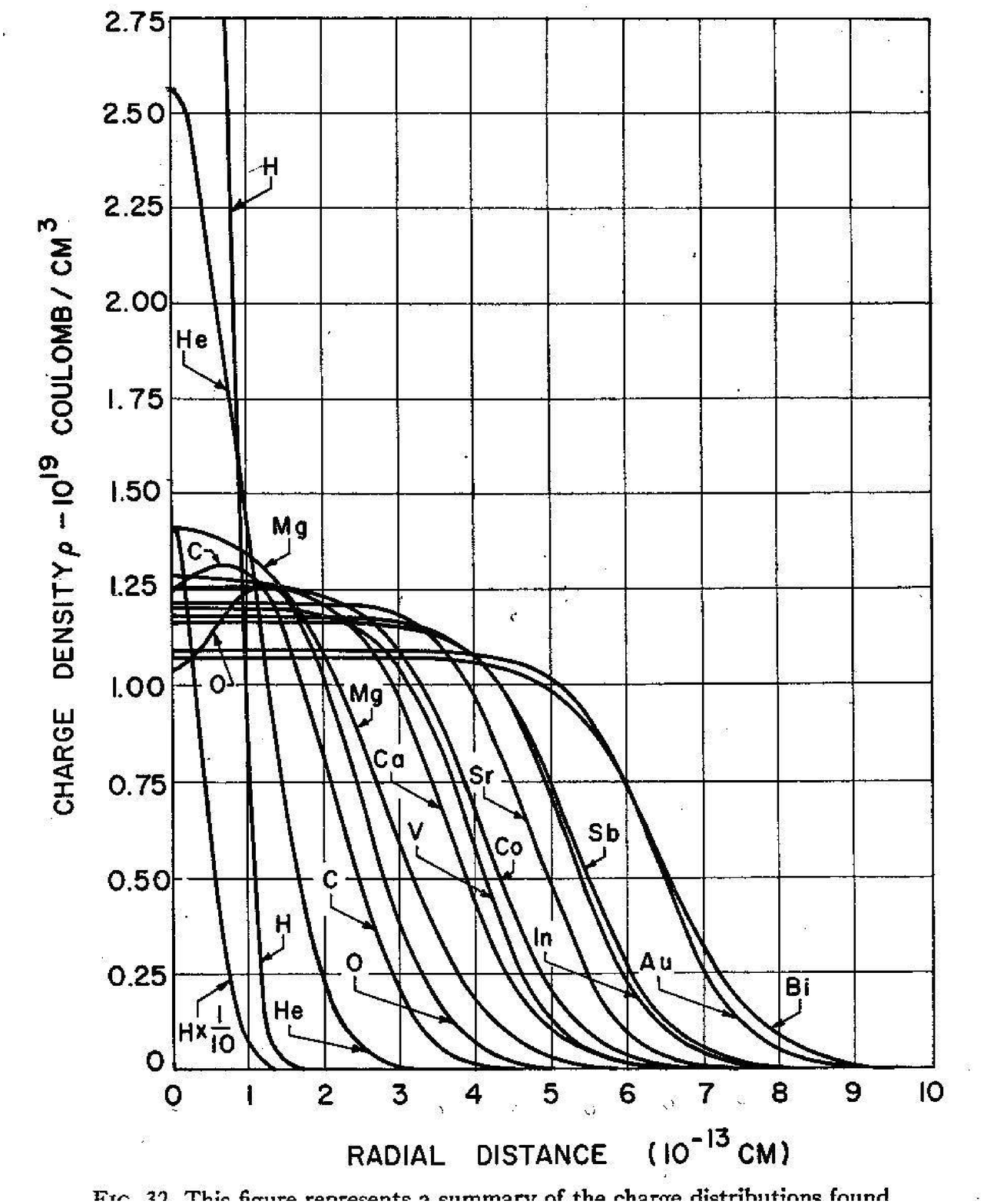}
\caption{Density distribution $H$, $^{4}\mbox{He}$ and other nuclei from ref.~\cite{w}}
\end{figure}
To understand this let us point out that the $t$- and $h$- nuclei are pretty much compact and have unique density distributions too as well. 

Let us look at their density distributions. We are plotting in Fig.1 and Fig.2 the density distribution of $\alpha$ and $^{3}\mbox{He}$~\cite{v} along with some standard nuclei~\cite{w}. One is struck by the remarkable fact that the 'average' central density of $\alpha$- and $^{3}\mbox{He}$ is about twice as large as the density of all the other nuclei. Note that the density of $t$- is also like that of $^{3}\mbox{He}$ nucleus~\cite{s}. Here the densities of $\alpha$-, $h$-, $t$- are completly different from those of other nuclei. This difference clearly indicates the fact that these $h$, $t$, and $\alpha$ should be treated differently from other nuclei. In addition $^{3}\mbox{He}$, $^{3}\mbox{H}$, and $^{4}\mbox{He}$ have a hole at the center which again marks them off as quite different from all other nuclei~\cite{s}. We suggest here that primary reason for the formations of clusters of $A=3$ and $A=4$ nuclei is due to their unique and identical density distributions. It also shows why no other nucleus may form good cluster substructures in nuclei. None have such high and hole-like density profiles!

\section{What is Alpha?}
Aa quoted from Langanke~\cite{n} $\alpha$- in the extreme shell model limit, can be treated equivalently as $(0S)^4$ or $\mbox{d+d}$ cluster or 3N+N cluster. But having shaken off this spurious encumberance in a previous section, we can see that $\alpha$- is not all that simple.

Langanke~\cite{n, p.97} has quoted the work of Kanada et al.~\cite{m} wherin they compare the binding energy for the ground state $(0^+_1)$ and the first excited state $(0^+_2)$ with different cluster model spaces. The calculation consider physical $p+{^{3}\mbox{H}}$, $n+{^{3}\mbox{He}}$, and $\mbox{d+d}$ cluster configurations plus also psuedoscalar excitations of deuterons. The conclusion is that as Langanke says, "{\it Contarary to a widely used picture, the $^{4}\mbox{He}$ ground state can not be well described as a bound state of two free deuterons. For a satisfactory description of $^{4}\mbox{He}$ ground state, the inclusion of $p+{^{3}\mbox{H}}$, $n+{^{3}\mbox{He}}$ configuration seems to be indispenseble. Also; both states have dominant $p+{^{3}\mbox{H}}$ and $n+{^{3}\mbox{He}}$ components.....}". 

What we are saying is that taking the above work seriously we feel that we are justified in taking, to a very good approximation, $^{4}\mbox{He}$ as made up of $p+{^{3}\mbox{H}}$ and $n+{^{3}\mbox{He}}$. So N=4 is actually 3+1 and not 1+1+1+1 or 2+2. This is quite amazing but consistent with all the experimental evidences, as shown above, supports this picture.

Alpha is unique not only because of its high binding energy of 28 MeV and high symmetry. Firstly, alpha has extremly high density and is about twice as great as that of the average nuclear matter. We have displayed this fact in Fig.1. and Fig.2. In addition $^{4}\mbox{He}$ has a hole at the centre of its density distribution which requires quark model to explain it~\cite{n} [s]. From the same figures also note the remarkable fact the density of $^{3}\mbox{He}$ is also equally twice as large as that of other nuclei. Both $^{4}\mbox{He}$ and $^{3}\mbox{He}$ are identical in this respect - i.e. extremely high central density. $^{3}\mbox{He}$ (also $^{3}\mbox{H}$) too have a 'hole'- depression in density distribution~\cite{s}. All $^{3}\mbox{He}$, $^{3}\mbox{H}$ and $^{4}\mbox{He}$ are similar in this regard - a property which no other nuclei have - twice as high density and a hole. 

Taking the work of Kanada et al.~\cite{m} and Langanke~\cite{n} seriously with the further improved understanding that there are no multicluster equivalences, we feel confident to state that we may justifiably write the $0^+_1$ and $0^+_2$ state wavefunctions of $^{4}\mbox{He}$ as follows
\begin{eqnarray}
\Psi_{0^+_1}&=&\frac{1}{\sqrt{2}}\left(\phi_h\times \psi_n - \psi_t\times \psi_p\right)\\
\Psi_{0^+_2}&=&\frac{1}{\sqrt{2}}\left(\phi_h\times \psi_n + \psi_t\times \psi_p\right)
\end{eqnarray}
where $\psi_h$ and $\psi_t$ represents the ground state wave functions of helion and triton respectively and with $\psi_n$, $\psi_p$ as wavefunctions of neutron and proton.

An experiment (in addition to the ones discussed above) which supports the picture of $\alpha$-being made up of a cluster of ($A=3$) is by Harwood and Kemper~\cite{x}. They found that 9.155, 9.83, and 10.69 MeV states in $^{15}\mbox{N}$ to have large cros sections in both three - and four particle transfer reactions from a study of $^{12}\mbox{C}({^{7}\mbox{Li}},\alpha\gamma)$,  $^{12}\mbox{C}({^{6}\mbox{Li}},{^{3}\mbox{He}}\gamma)$ and  $^{11}\mbox{B}({^{7}\mbox{Li}},t\gamma)$ reactions at 28, 34 and 28 MeV respectively. They found that these same states are populated strongly in both the three - and four - particle transfer reactions. They suggested that this shows that there is a large overlap between the triton and $\alpha$- cluster states. Theoretical models were unable to fit this properly. Now, why should $\alpha$- and $t$- transfer reactions should give above identical results. In our picture it is easy to understand this. As $\alpha$- is a nucleus with $t+p$ cluster, thus the smilarity of the above result shows that both in the case of $t$- transfer and for $\alpha$-transfer its the $t$- constituent, which is producing the same spectra. Hence their experiment supports our model of alpha. We clarify this point as follows. As per cluster $^{15}_7\mbox{N}_8={^{12}_6\mbox{C}_6}+{^{3}_1\mbox{H}_2}$ excites the observed states in $t$-transfer. Similarly $^{15}_7\mbox{N}_8={^{11}_5\mbox{B}_6}+{^{4}_2\mbox{He}_2}$ excites the states observed in $\alpha$-transfer. Now these states which are common to both can be understood if as we suggested, alpha has a structure of $\alpha\sim t+p$. Then if the common state has the structure of $^{12}\mbox{C}+t$ in which case $^{11}\mbox{B}+\alpha\rightarrow {^{12}\mbox{C}}+t$. If the common state has structure $^{11}\mbox{B}+\alpha$ then in the first case $^{12}\mbox{C}+t\rightarrow {^{11}\mbox{B}}+\alpha$. Thus this supports our picture of $\alpha$-nucleus being formed of ($A=3$) cluster as basic entity. 

Alpha clustering in nuclei is well known and a well studied problem. In as much as this is true the above wavefunction of alpha for the ground state and the first excited state, should be taken as the most straightforward proof of $A=3$, triton and helion clustering in nuclei. If alphas are real in nuclei then so must triton and helion be as per the above equation. In simple words, $\alpha$-cluster is itself made up of $A=3$ clusters.

\section{Towards Our Model}
We look at two authors who historically were the first ones to include $A=3$ clusters systematically include. First one was Pauling~\cite{y} where he assumed spherons as building blocks of all nuclei. A spheron is a cluster which forms in localized $0S$ orbitals. These are $nn$, deuteron, helion, triton and $\alpha$- clusters. He tried to reproduce magic structures using spherons. He arranged these spherons geometrically into closed packed symmetrical structures to obtain magic numbers. The idea of spherons as building blocks with simple clusters (in addition to $n$, $p$ as building blocks) was interesting and was reasonably successful as a starting model~\cite{a}.

The other model is that of Mac Gregor~\cite{b}. He used $n$, $p$, $h$, $t$, and $\alpha$ as building blocks to reproduce nucleon binding energies. ''{\it For its very large binding energy, the $\alpha$-particle appears in some sense, as a ''fully saturated nuclei''. Hence it seems reasonable to consider clusters as constituting a complete basis set for reproducing atomic nuclei}''~\cite{b}. He assumed a novel two dimensional Ising model for nuclei. He too had partial successes~\cite{a}.

Here, while not pursuing on the lines of their respective models - that of close packing of Pauling~\cite{v} and Ising Model for Mac Gregor~\cite{b} we give them credit for recognizing that $A=3$ nuclei - $h$ and $t$ be condidered to be playing a more fundamental role in understanding the atomic nucleus.

In terms of matter density distribution we notice that there are three blockings as per Fig.1 and Fig.2. These having $n$ and $p$ as one block, $h$, $t$ and $\alpha$ as another block and the third one as that of all the other nuclei. We know that historically the similarity recognized from the similarity of masses of $n$ and $p$ (but here clearly apparent in terms of matter distribution also) was recognized as leading to that existence of an isospin symmetry $SU(2)_T$ where fundamental representation is $(n,p)$. In the third set all the other nuclei have similar pattern of matter distribution. What is the significance of similarity of matter distribution of $h$, $t$, and $\alpha$? Note that this similarity is more deep as only these three have, as pointed above, holes at the centre of their density distribution as well. Such a hole is not found in the first block, that of $(n,p)$ or in the third block, that of all the other nuclei.

Noting this categorization we can now appreciate now Mac Gregor's classification of assuming $n$, $p$, $h$, $t$, and $\alpha$ as building blocks of nuclei~\cite{b}. Meaning in his model $n$, $p$, $h$, $t$, and $\alpha$ are not treated as nuclei, but from these are treated basic building blocks from which all the other nuclei are made up of these only. This is similar to how quarks are building blocks of proton and neutron.

Hence pointing out that density distribution of $(n,p)$ and of ($h$, $t$, and $\alpha$) are giving unambiguos support of this concept of building blocks, we agree with Mac Gregor's classification~\cite{b}. However, we go beyond his concept, in pointing out that within ($h$, $t$, and $\alpha$) blocking, there is actually another intrinsic blocking, that of $(h, t)$ as being different from $\alpha$-nucleus.

It has been shown that actually just like $(n, p)$ forms a block and this manifests itself by inducing the nuclear $SU(2)_T$ isospin group, $(h, t)$ also forms a block and thereby induces a new $SU(2)_{\cal A}$ symmetry group the so called nusospin group~\cite{z}. Thus the fundamental block of nuclei as per this picture are actually $(n, p)$ and $(h, t)$ only. 

Then what is $\alpha$ particle? As we have discussed the $\alpha$-particle is actually the first and basic saturated nucleus predominantly have the structure 3+1, and hence that its made up of $(h, t)$ and $(n, p)$ fundamental entities~\cite{aa}.

There are several evidences that $(n, p)$ exist as fundamental entities in nuclei. This fact, has been so dominant in our present understanding of nuclear physics, that it has prevented physicists from recognizing the presence of $(h, t)$ as fundamental entities too. In spite of our gut feeling, surprisingly there have been a large number of experimental evidences of the presence of $h$ and $t$ in nuclei as we have discussed and highlighted these in the previous sections. Because of antisymmetrization with respect to nucleons degrees in nuclei the possibility of having pre-existing $h$ and $t$ appears to be negligible. But they are there! By and large people have tended to accept the presence of $\alpha$'s in nuclei and hence that does not constitute a problem to most. However presence of $h$ and $t$ inside nuclei as shown by experiments discussed here should be treated as an anomoly. 

In our model the basic building blocks are only four $n$, $p$, $t$ and $h$. We do not consider deuteron as a building block as it is made up of $n$ and $p$ which themselves are building blocks. As per our definition, our basic building block entity should not be considered as being made up of other basic building entities. For the same reason, we also do not consider $\alpha$-particle as a basic building block, because as we have seen, empirically there are convincing evidences that it is made up of $A=3$ cluster plus one nucleon. So there is a $t$ and $h$ substructure in $\alpha$. We treat $\alpha$- as the first and most basic saturated nucleus (note its BE/A is 7 MeV very close to 8 MeV/A for heavy nuclei). Alpha keeps popping up as a cluster in the build up of heavier nuclei - but not as a basic building block but as a good cluster which nuclear dynamics prefers to create. The reason here for rejecting $\alpha$- as a basic building block is the same as that of rejecting $d$- as a building block - both are built up of other basic building blocks viz: $n$, $p$, $h$ and $t$. Note that it is for the same reason that $^{6}\mbox{Li}$ as having a structure $h+t$ is not a building block too.

In the case of $h$ and $t$, though the binding energy is only $\sim$8 MeV but these are unique entities in that both $h$ and $t$ are {\underline{only nuclei}} known (beside deuteron) which have no excited states~\cite{bb}. So within the nuclear medium as long as relative excitation energies are small then these can be treated as basic building blocks~\cite{q,z,aa,bb}.

There are additional reasons due to density distribution of these nuclei (as discussed above) which make them into good basic building blocks of nuclei. But again it is experiments - which we have already discussed, which keep on observing that $h$- and $t$- both exist as primary building blocks inside nuclei. Again we refer to Mac Gregor's~\cite{b} extensive discussion of various empirical evidences including presence of $h$- and $t$- (along with $\alpha$) as preformed clusters deep inside nuclei.

One may ask- but we know that $h$ and $t$ are made up of $n$ and $p$ and so how is it justified to take them up as a building block. the answer broadly is similar to the reason as to why $p$ and $n$ though being made up of three quarks can still be considered as building blocks for nuclear physics.

The reason for both the above situation (i.e. $n/p$ made up of 3 quarks and $h/t$ made up of three nucleons but still basic) is fundamentally empirical - it is experiments which is telling us so. It is from experiment that we know that upto about 300 MeV excitation the three quarks primarily remains bound in a proton and neutron structure. Similarly it is experiment which show that $h$ and $t$ can be treated as fundamental (in specific situations). Just as it is due to specific 3BF (in quark context) manifested through colour confinement in QCD for three quarks which makes $SU(2)_I$ for ($u,d$) quarks to go over to $SU(2)_I$ for ($n,p$) as a new nuclear isospin group. In the same manner it is specific nuclear 3BF in $A=3$ nuclei which make them unique and different, and thus $SU(2)_I$ of ($n,p$) induces a new symmetry, $SU(2)_{\cal A}$, the so called nusospin~\cite{z,aa} where ($h, t$) appears as a fundamental entity. Thus the complete group structure in nuclei is $SU(2)_T\times SU(2)_{\cal A}$. The other reasons which led to the introduction of nusospin group $SU(2)_{\cal A}$ in the first place has already been made and the reader is reffered to them ~\cite{z,aa}. In this paper we have included further independent reasons which have led us to suggest $h$, $t$ as basic building blocks and thus leading to the same symmetry structure of $SU(2)_{\cal A}$. We may note that the wave function of $0^+_1$ and $0^+_2$ of $\alpha$-particle as given above (Eq.2 and 3) was also suggested earlier on the basis of $SU(2)_{\cal A}$ nusospin group arguments ~\cite{aa}. 

\section{Further support for $SU(2)_{\cal A}$ Nusospin Symmetry}
Nagatani et al.~\cite{cc} have demonstrated high selectivity in reactions involving transfer of three nucleons which is just as good as was the case for $\alpha$- particle transfer and also transfers to mirror final states yield essentially identical spectra. Most remarkably, the same selective population of certain state was essentially the same under the change of energy or reaction. Very significant is the fact that essentially identical spectra was found in all the helion transfer studies by different groups: $^{12}\mbox{C}({^{6}\mbox{Li}},t){^{15}\mbox{O}}$; $^{12}\mbox{C}({^{11}\mbox{B}},{^{8}\mbox{Li}}){^{15}\mbox{O}}$, $^{12}\mbox{C}({^{12}\mbox{C}},{^{9}\mbox{Be}}){^{15}\mbox{O}}$~\cite{dd,ee}. Hence they concluded~\cite{cc} that these reactions represent a powerful reaction-mechanism-independent method of finding corresponding analogue states in the residual nuclei.

The significance of analog states for mirror nuclei is due to isospin symmetry $SU(2)_T$ of ($n,p$) fundamental representation. Isospin invariance means that the wave function of a given isospin T are unchanged if we replace some of the neutrons by protons or vice versa. This transformation takes us from one member of isobar to another. The wave function of these two isobaric analogue states are related to each other through isospin raising and lowering operator. Hence the two states must have essentially the same properties except that some of the neutrons are replaced by corresponding protons. A special case is that of mirror nuclei which have the same number of nucleons except the mirror change of number of protons and neutrons. Thus isospin requires that in these mirror nuclei, the energy level spectra and the properties of various states should be  similar (ignoring coulumb interaction) to each other. But are the states being populated in these mirror nuclei in the above experiments are are due to the nuclear $SU(2)_T$ isospin symmetry requirements? To understand the issue we discuss as follows.

Suppose we were not aware of isospin symmetry $SU(2)_T$ between $n$ and $p$ states and we did same experiments of a single nucleon transfer only and found spectra independent of whether it was $n$ or $p$ which was transfered. So for example for $^{15}\mbox{O}$ and $^{15}\mbox{N}$ we may have the structure
\begin{eqnarray*}
{^{15}_8\mbox{O}_7}\sim {^{14}_7\mbox{N}_7}+p\\
\end{eqnarray*}
and
\begin{eqnarray*}
{^{15}_7\mbox{N}_8}\sim {^{14}_7\mbox{N}_7}+n
\end{eqnarray*}
and suppose that in these residual nuclei, in one nucleon transfer reaction we found similar spectra, then we would conclude that there is a ''new symmetry'' indicating invariance under the change $n\leftrightarrow p$. And that this would be $SU(2)_T$ the isospin symmetry for ($n,p$) representation.

But here in the experiments under discussion it is {\bf one shot transfer} of $h$- and $t$- as a single entity. The residual nuclei which show identity of spectra have actually a structure
\begin{eqnarray*}
{^{15}_8\mbox{O}_7}\sim {^{16}_6\mbox{C}_6}+{^{3}_2\mbox{He}_1}\\
\end{eqnarray*}
and
\begin{eqnarray*}
{^{15}_7\mbox{N}_8}\sim {^{12}_6\mbox{C}_6}+{^{3}_1\mbox{H}_2}
\end{eqnarray*}
hence the identity of spectra here should be indicature of a new symmetry where the spectra does not distinguish between the exchange of ${^{3}_2\mbox{He}_1}\leftrightarrow {^{3}_1\mbox{H}_2}$. this means that there should exist a 'new' $SU(2)$ symmetry with ($h,t$) forming the fundamental representation. And indeed this is exactly the $SU(2)_{\cal A}$ nusospin symmetry already discussed. Hence, in contrast to what they claimed~\cite{cc}, these experiments are demonstrating invariance under the new nusospin symmetry $SU(2)_{\cal A}$ and not under the $SU(2)_T$ isospin symmetry. This clearly shows that the ($h,t$) as leading to a new $SU(2)_{\cal A}$ nusospin symmetry be taken as fundamental, actually fundamental as the $SU(2)_T$ ($n,p$) isospin symmetry.

\section{conlcusion}
We have presented here results from a large number of experiments which point to the presence of $A=3$ i.e. triton and helion clusters, in nuclei. Some of the experimental evidence presented here give very compelling support for preformed and existing $A=3$ clusters in nuclei. Most of the other experiments do give reasonably good support to the above concept. Taken together, all these experiments, give strong reasons to believe in the existence of tritons and helions in nuclei (along with the alpha as well). But this picture is in conflict with what may be said to constitute the generally accepted picture of nuclear physics today. We have discussed the reasons for this opposition and as to how rectify the situations so that one is able to appreciate both the presence of $t$- and $h$- clusters in nuclei (most having already reconciled themselves to the $\alpha$-cluster though). this allows us to provide a more unified (and consistent with all the above experiments) description of nuclei where triton and helion plays a more fundamental role.


\end{document}